# Emergence of Two-Dimensional Massless Dirac Fermions, Chiral Pseudospins, and Berry's Phase in Potassium Doped Few-Layer Black Phosphorus

*Seung Su Baik,[1,2] Keun Su Kim,[3,4] Yeonjin Yi,[1] and Hyoung Joon Choi[1,2],\**

[1] Department of Physics, Yonsei University, Seoul 120-749, Korea.

[2] Center for Computational Studies of Advanced Electronic Material Properties, Yonsei University, Seoul 120-749, Korea.

[3] Department of Physics, Pohang University of Science and Technology, Pohang 790-784, Korea.

[4] Center for Artificial Low Dimensional Electronic Systems, Institute for Basic Science, Pohang 790-784, Korea.

**ABSTRACT:** Thin flakes of black phosphorus (BP) are a two-dimensional (2D) semiconductor whose energy gap is predicted being sensitive to the number of layers and external perturbations. Very recently, it was found that a simple method of potassium (K) doping on the surface of BP closes its band gap completely, producing a Dirac semimetal state with a linear band dispersion in the armchair direction and a quadratic one in the zigzag direction. Here, based on first-principles density functional calculations, we predict that, beyond the critical K density of the gap closure, 2D massless Dirac Fermions (i.e., Dirac cones) emerge in K-doped few-layer BP, with linear band dispersions in all momentum directions, and the electronic states around Dirac points have chiral pseudospins and Berry's phase. These features are robust with respect to the spin-orbit interaction and may lead to graphene-like electronic transport properties with greater flexibility for potential device applications.

Black phosphorus (BP)[1,2] is a layered material consisting of stacks of a wrinkled honeycomb lattice with single phosphorus elements. In its bulk state, pristine BP has a direct energy gap of 0.33 eV,[3,4] and mechanically exfoliated thin flakes of BP[5-9] have an increasingly larger energy gap as the BP thickness decreases.[10-12] These energy gaps of BP depend sensitively on external perturbations such as strain,[13-15] pressure,[16] and electric fields.[15,17-20] Previous theoretical studies showed that few-layer BP can be switched to a topologically nontrivial insulator by external electric field of ~$10^9$ V/m which causes an inversion of the energy gap,[20] and bulk BP can possess Dirac cones if the energy gap is inverted by sufficient pressure.[16] Very recently, it was found that a simple method of potassium (K) doping on the surface of BP closes its band gap completely,[21] producing a Dirac semimetal state with a linear dispersion in the armchair direction and a quadratic one in the zigzag direction.[21] In the present work, we consider potassium doping as a method to invert the energy gap of few-layer BP beyond the gap closing, and study its electronic structure using first-principles density functional calculations. We predict an emergence of 2D massless Dirac cones with chiral pseudospins and Berry's phase in K doped few-layer BP. The 2D Dirac cones have linear band dispersions in all momentum directions and they are robust with respect to the spin-orbit interaction. The spin-orbit interaction lifts the spin degeneracy but it does not open a gap at the Dirac points, resulting in spin-polarized Dirac cones separated slightly in the momentum space. Since the energy gap and the band inversion can be controlled by the potassium doping, the 2D massless Dirac cones in the band-inverted BP may open a new pathway for high performance devices.

We consider a four-layer BP as a prototype of a few-layer BP as shown in Figure 1a, where BP has a puckered structure with the zigzag direction along the $x$ axis and the armchair direction along the $y$ axis.[2] We performed first-principles density functional calculations of a four-layer BP using the SIESTA code.[22] The generalized gradient approximation (GGA)[23] was employed for the exchange correlation. Details of calculations are described in the Computational Methods section.

Previous studies of BP have shown that the conduction and valence band dispersions show quadratic dependence on *k* along the zigzag direction and asymptotic linear one along the armchair direction.[7,8] Our calculated results of pristine four-layer BP before the deposition of the potassium atoms confirm these features with an energy gap of 0.36 eV, as shown in Figure 1b and 1c. The conduction-band effective mass from our calculation is 1.34 $m_e$ in the zigzag direction and 0.13 $m_e$ in the armchair direction, and the valence-band one is 1.49 $m_e$ in the zigzag direction and 0.12 $m_e$ in the armchair direction. Here $m_e$ is the electron mass in vacuum. The effective Hamiltonian of the conduction and valence bands can be given by

$$H = \begin{pmatrix} 0 & \frac{1}{2}E_g + \frac{\hbar^2 k_x^2}{2m^*} - i\hbar v_y k_y \\ \frac{1}{2}E_g + \frac{\hbar^2 k_x^2}{2m^*} + i\hbar v_y k_y & 0 \end{pmatrix} = \left(\frac{1}{2}E_g + \frac{\hbar^2 k_x^2}{2m^*}\right)\sigma_x + \hbar v_y k_y \sigma_y, \quad (1)$$

where $E_g$ = 0.36 eV is the energy gap, $\hbar$ is the Planck's constant divided by $2\pi$, $m^*$ = 1.42 $m_e$ is the average effective mass of the two bands along the zigzag direction, $v_y$ = 5.6×10$^5$ m/s is the velocity along the armchair direction, and $\sigma_x = \begin{pmatrix} 0 & 1 \\ 1 & 0 \end{pmatrix}$ and $\sigma_y = \begin{pmatrix} 0 & -i \\ i & 0 \end{pmatrix}$ are Pauli matrices. In the Hamiltonian (1), we neglect the difference in the effective mass for the conduction and valence bands for simplicity.

When potassium (K) atoms are deposited on top of BP as shown in Figure 2a, the band gap decreases rapidly due to the giant Stark effect.[21] When we introduce one K atom per 2×2 surface unit cell of four-layer BP, which corresponds to the areal density of the deposited K atoms of 1.65×10$^{14}$ /cm$^2$, and relax the positions of the K atom and P atoms in the topmost BP layer, our band calculation shows that the valence band and the conduction band are inverted at the Γ point with a negative value for $E_g$, and the two bands cross with each other at two $k_x$ points, which are marked with $k_x = k_D$ and $k_x = -k_D$ in the figure. Very interestingly, the two crossing points are, in fact, two Dirac points. Our calculated band dispersions along the $k_y$ direction at the two crossing points are linear as shown in

Figure 2d. The conduction and valence bands have linear dispersion at $(k_x, k_y) = (\pm k_D, 0)$ regardless of the k-vector direction. Now, the effective Hamiltonian can be expressed as

$$H = \begin{pmatrix} 0 & \frac{\hbar^2}{2m^*}(k_x^2 - k_D^2) - i\hbar v_y k_y \\ \frac{\hbar^2}{2m^*}(k_x^2 - k_D^2) + i\hbar v_y k_y & 0 \end{pmatrix} = \frac{\hbar^2}{2m^*}(k_x^2 - k_D^2)\sigma_x + \hbar v_y k_y \sigma_y, \quad (2)$$

where the negative energy gap $E_g$ at the $\Gamma$ point and the position $k_D$ of the Dirac point are related by $E_g = -\hbar^2 k_D^2/m^*$. Near the Dirac point at $k_x = k_D$, this Hamiltonian reduces to

$$H = \begin{pmatrix} 0 & \hbar v_x k_x' - i\hbar v_y k_y \\ \hbar v_x k_x' + i\hbar v_y k_y & 0 \end{pmatrix} = \hbar v_x k_x' \sigma_x + \hbar v_y k_y \sigma_y, \quad (3)$$

where $v_x = \hbar k_D/m^* = 0.86 \times 10^5$ m/s is the velocity along the zigzag direction at the Dirac point and $k_x' = k_x - k_D$ is the value of $k_x$ with respect to the Dirac point. Here $v_y = 2.8 \times 10^5$ m/s. The Hamiltonian (3) implies that the chiral pseudospin should exist, which we can confirm by calculating the electronic wave functions near the Dirac point. As shown in Figure 3, electronic wave functions at an energy slightly above the Dirac point change rapidly depending on the k-vector direction. We can assign spinor representations $(\frac{1}{\sqrt{2}}, \frac{1}{\sqrt{2}})$, $(\frac{1}{\sqrt{2}}, \frac{i}{\sqrt{2}})$, $(\frac{1}{\sqrt{2}}, \frac{-1}{\sqrt{2}})$, and $(\frac{1}{\sqrt{2}}, \frac{-i}{\sqrt{2}})$ for the wave functions at 0°, 90°, 180°, and 270°, respectively. Then the azimuthal angle $\theta$ of the spinor representation $(\frac{1}{\sqrt{2}}, \frac{e^{i\theta}}{\sqrt{2}})$ changes from 0 to $2\pi$ around the Dirac point, resulting in Berry's phase of $\pi$. Due to 2D massless Dirac Fermions, chiral pseudospins, and Berry's phase, gap-inverted few-layer BPs have strong potential for extremely high mobility, anomalous quantum Hall effect, and other exotic phenomena, similarly to graphene.[24-29]

So far we did not consider the spin-orbit interaction in the K-doped BP system. When the spin-orbit interaction is included in our calculation using additional Kleinman-Bylander projectors derived from fully relativistic j-dependent pseudopotentials,[30] the two Dirac points are split into four points as shown in Figure 2e, moving from $(k_x, k_y) = (\pm k_D, 0)$ to $(k_x, k_y) = (\pm k_D, \pm k_{SO})$ with no energy gap at the four new Dirac points. The spin degeneracy is now lifted so that the electron spins of the upper and lower Dirac cones at $(k_x, k_y) = (\pm k_D, k_{SO})$ are almost aligned in the +x direction while those at $(k_x, k_y) = (\pm k_D, -k_{SO})$ are almost in the -x direction, with the spin x component being 0.99~0.98 times $\hbar/2$ and the spin y component being only 0.00~0.20 times $\hbar/2$. This implies that the spin-orbit interaction near Dirac points is approximately a product of the real-spin x component and the pseudospin y component, $H_{SO} = \lambda S_x \sigma_y$, where $\lambda$ is a constant. The value of $k_{SO}$ is $1.5 \times 10^7$ /m so that the conduction and valence bands have an energy difference of $2\hbar k_{SO} v_y = 5$ meV at the old Dirac points $(k_x, k_y) = (\pm k_D, 0)$ despite of no energy gap at the new Dirac points $(k_x, k_y) = (\pm k_D, \pm k_{SO})$.

Finally, we discuss the crossover from the 2D Dirac semimetal to the insulator. As the amount of deposited K atoms reduces, the magnitude of the inverted energy gap $|E_g|$ decreases and the Dirac points move toward the Γ point. In our calculations, we can reduce the effect of the K atoms gradually by moving them farther away from their equilibrium distance from BP, which is equivalent to reducing the density of deposited K atoms experimentally. Thus, at a critical density of the K atoms, the valence band and the conducting band touch only at the Γ point,[21] and the two bands are quadratic along the x direction and linear along the y direction,[21] as shown in Figure 4. We can write the low-energy effective Hamiltonian at this critical point as

$$H = \begin{pmatrix} 0 & \frac{\hbar^2 k_x^2}{2m^*} - i\hbar v_y k_y \\ \frac{\hbar^2 k_x^2}{2m^*} + i\hbar v_y k_y & 0 \end{pmatrix} = \frac{\hbar^2 k_x^2}{2m^*} \sigma_x + \hbar v_y k_y \sigma_y, \qquad (4)$$

which is the same as the Hamiltonian (2) with $k_D = 0$ and also the same as the Hamiltonian (1) with $E_g = 0$. Here $v_y = 3.0 \times 10^5$ m/s. The Hamiltonian (4) shows a limited behavior of chiral pseudospins. When we consider the wave functions around the Dirac point at an energy slightly above the Dirac point, the azimuthal angle $\theta$ of the spinor representation $(\frac{1}{\sqrt{2}}, \frac{e^{i\theta}}{\sqrt{2}})$ of the wave function increases from 0 to π/2, then decreases from π/2 to -π/2, and eventually increases from -π/2 to 0, resulting in zero Berry's phase.

Our supercell calculations with K-doped BPs (Figures S4 and S5 in Supporting Information) show a linear dependence of the band-gap reduction on the K doping density with a rate of 0.28 eV per $1 \times 10^{14}$/cm$^2$ and the critical density of $1.3 \times 10^{14}$/cm$^2$ for the band-gap closing. In our calculation, the band gap of the pristine four-layer BP is about 0.3 eV smaller than the hybrid functional calculations (HSE06).[18,20] Thus, when we consider this 0.3 eV difference, the critical K doping density changes to $2.4 \times 10^{14}$/cm$^2$. Meanwhile, since the energy gap decreases monotonically with the increase of the BP thickness,[8,11,12,18,20] the band inversion and thereby the emergence of 2D Dirac cones occur at smaller critical K densities in thicker BP layers.

In conclusion, we have introduced K atoms on the surface of four-layer BP to invert the band gap of pristine BP. As the band inversion occurs, the overlapped bands make only two crossing points in the momentum space and 2D massless Dirac Fermions emerge at those points, entailing the chiral pseudospins and the Berry's phase. We have shown that the emerged Dirac cones are robust with respect to the spin-orbit interaction. Without opening a gap, the spin-orbit coupling slightly shifts the spin-degenerate Dirac cones in opposite directions depending on the real-spin polarization direction. Because the band inversion can be controlled by the surface density of K atoms, the results presented here are the switchable massless Dirac Fermions in a 2D material, which would be applied for the new device developments.

**Computational Methods**

We first determined the equilibrium lattice constants and internal parameters of bulk BP (space group number: 64, Bmab) using the WIEN2k code[31] and obtained the equilibrium lattice constants of a = 3.32, b = 4.57, and c = 11.33 Å. We used the Perdew-Burke-Ernzerhof-type (PBE) generalized gradient approximation (GGA)[23] for the exchange-correlation. The calculated band gaps ($E_g$) are 0.21 eV for bulk BP and 0.98 eV for phosphorene (monolayer BP). These values agree well with the previous experimental and theoretical reports.[7,8] The optimization of cell parameters and atomic positions is crucial in achieving the finite band gap without hybrid functionals or GW approximations.[10] As shown in Figure S1, we confirmed that the overall band structure with GGA is in good agreement with experimental ARPES data and the hybrid functional results.[7] Then, using the optimized bulk structure, we constructed multilayer BP structures and calculated their band structures using the SIESTA code,[22] where the exchange-correlation was treated with GGA (PBE). The electronic wave functions and charge densities were projected onto a real-space grid with an equivalent energy cutoff of 500 Ry and integrated using a 12×12 k-grid in the 2D Brillouin zone for all multilayer-BP calculations. The spin-orbit interaction is incorporated within a fully relativistic $j$-dependent pseudopotential[32] and treated in the $l$-dependent fully separable nonlocal form using additional Kleinman-Bylander projectors.[33,34] Our method for the spin-orbit interaction is well tested with the Rashba splitting of the Au (111) surface states[30] and topological surface states of $Bi_2Se_3$.[35]

For K-doped BP, a single K atom was added on each 2×2 surface unit cell of multilayer BP. Atomic relaxation was performed for the K atom and the topmost P layer, and atomic displacements of P atoms were found smaller than 0.1 Å. In order to describe a lower K density effectively, the vertical distance between K and BP is gradually increased from the equilibrium (2.7 Å), which reduces the effect of the K atom without change in the supercell size. The validity of this method was checked by comparison with results from larger supercell calculations (up to 4×4 surface supercell and 6-layer BP) with lower K densities at equilibrium distance.

Figures

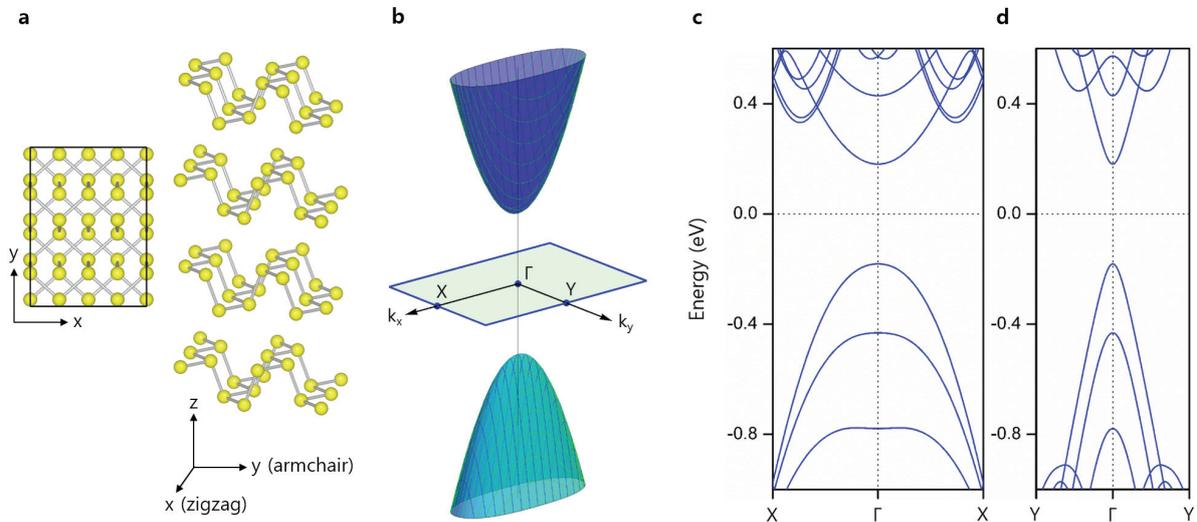

**Figure 1.** Pristine four-layer BP. (a) Top and side views of the atomic structure. Yellow dots are P atoms. (b) $k_x$-$k_y$ plot of calculated conduction and valence bands. (c) Calculated band structure along the $k_x$ axis (the zigzag direction). (d) Calculated band structure along the $k_y$ axis (the armchair direction).

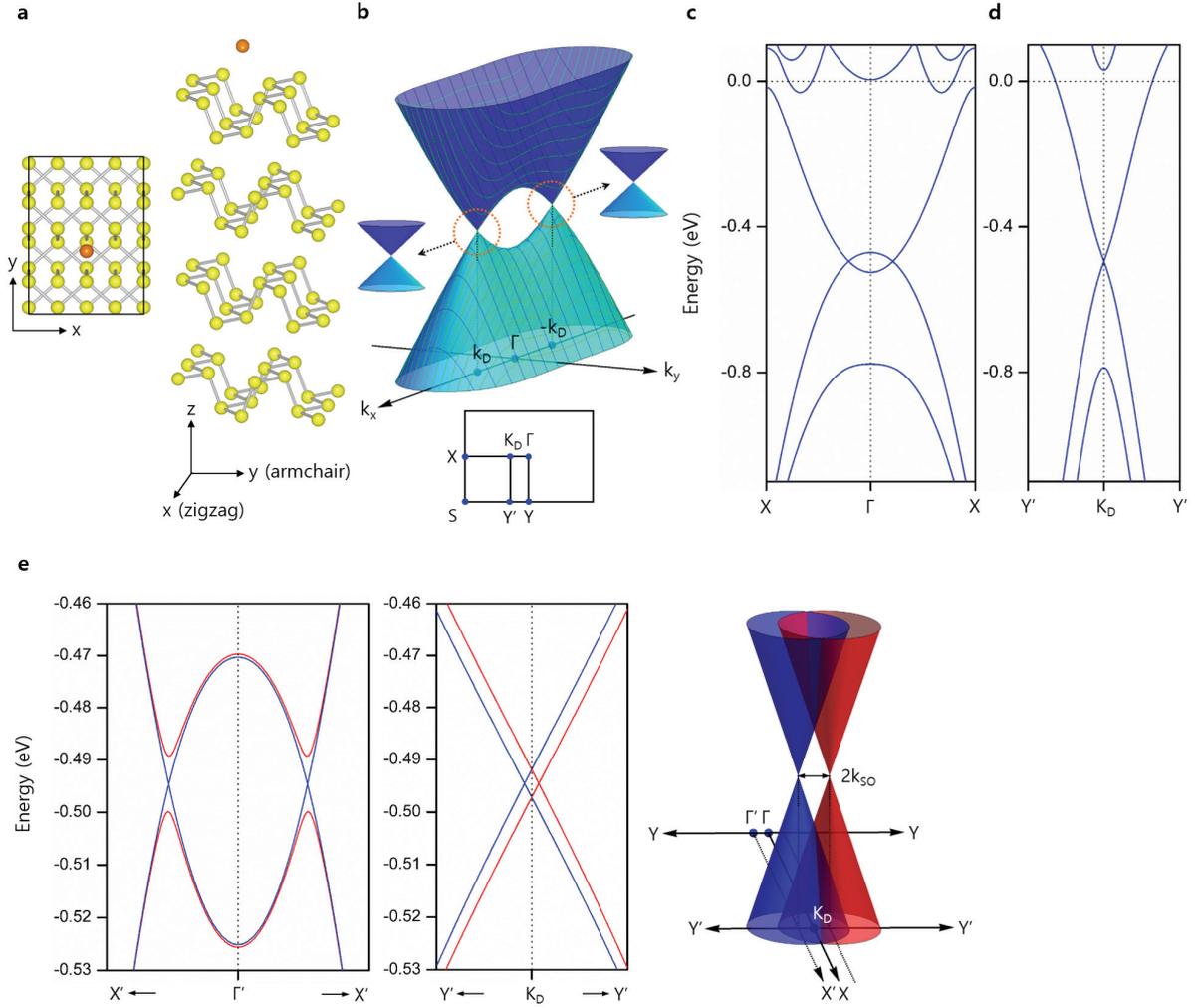

**Figure 2.** K-doped four-layer BP. (a) Top and side views of atomic structure. Yellow dots are P atoms and the orange one is the K atom. The K atom is 2.7 Å away from the topmost P layer. (b) $k_x$-$k_y$ plot of calculated conduction and valence bands. (c) Calculated band structure along the $k_x$ axis (the zigzag direction). (d) Calculated band structure along the $k_y'$ axis (the armchair direction at one of the Dirac points). The $k_y'$ axis is in parallel with the $k_y$ axis, passing through $(k_D, 0)$. (e) Calculated band structures including the spin-orbit interaction, plotted along the $k_x'$ and $k_y'$ axes. The $k_x'$ axis is in parallel with the $k_x$ axis, passing through $(\pm k_D, -k_{SO})$. Red lines represent bands with spins almost aligned in the $+x$ direction while blue ones in the $-x$ direction. With the spin-orbit interaction, each Dirac cone at $(\pm k_D, 0)$ is split into two spin-resolved cones which are separated by $2k_{SO}$ along the $k_y'$ axis. Spins of the upper and lower Dirac cones at $(k_x, k_y) = (\pm k_D, k_{SO})$ are almost aligned in the $+x$

direction, colored in red, while those at $(k_x, k_y) = (\pm k_D, -k_{SO})$ are almost in the -*x* direction, colored in blue.

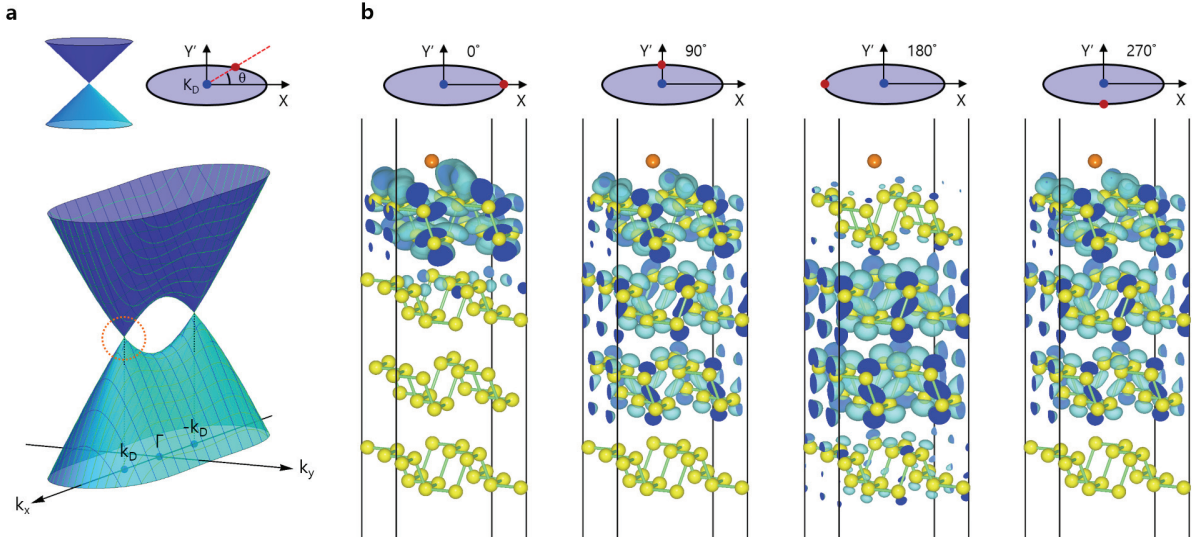

**Figure 3.** Chiral pseudospin feature of the K-doped BP. (a) Band structures near Dirac points. (b) Electronic wave functions around the Dirac point at $k_x = k_D$. Wave functions are plotted at an energy slightly above the Dirac point, with an isosurface level set to $2.5 \times 10^{-3}/\text{Å}^3$. We can assign spinor representations $(\frac{1}{\sqrt{2}}, \frac{1}{\sqrt{2}})$, $(\frac{1}{\sqrt{2}}, \frac{i}{\sqrt{2}})$, $(\frac{1}{\sqrt{2}}, \frac{-1}{\sqrt{2}})$, and $(\frac{1}{\sqrt{2}}, \frac{-i}{\sqrt{2}})$ for the wave functions at 0˚, 90˚, 180˚, and 270˚, respectively. Here, the $(\frac{1}{\sqrt{2}}, \frac{1}{\sqrt{2}})$ state, which is localized mainly in the topmost BP layer, is close to the conduction-band-minimum state just before the band-gap closing, and the $(\frac{1}{\sqrt{2}}, \frac{-1}{\sqrt{2}})$ state, which is away from the topmost BP layer, is to the valence-band-maximum state. The $(\frac{1}{\sqrt{2}}, \frac{i}{\sqrt{2}})$ and $(\frac{1}{\sqrt{2}}, \frac{-i}{\sqrt{2}})$ states are linear combinations of $(\frac{1}{\sqrt{2}}, \frac{1}{\sqrt{2}})$ and $(\frac{1}{\sqrt{2}}, \frac{-1}{\sqrt{2}})$ so that the states are spread throughout the topmost to the third BP layer.

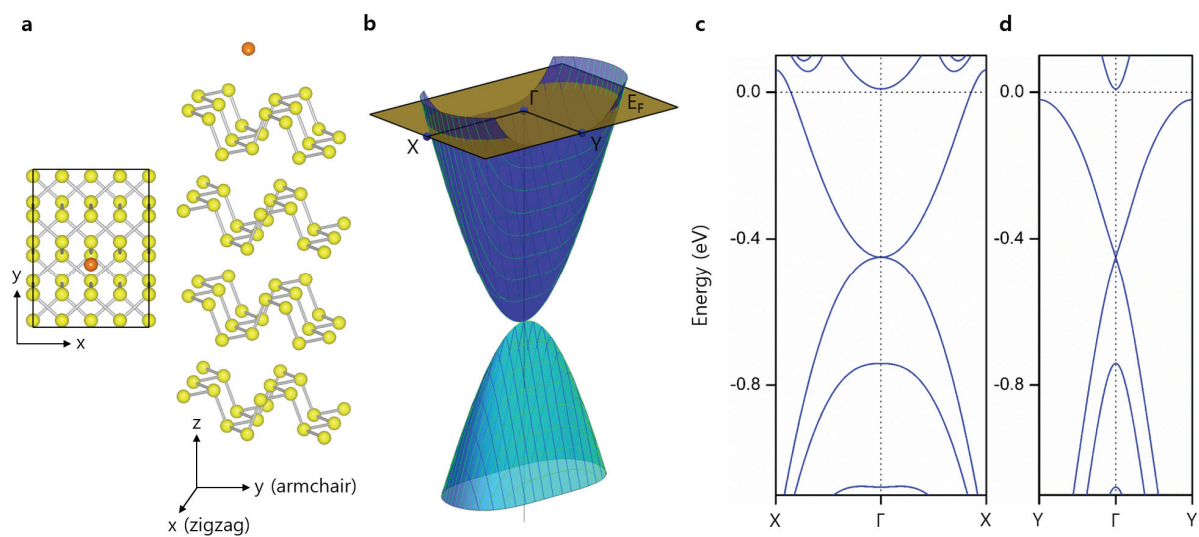

**Figure 4.** Critically K-doped four-layer BP. (a) Top and side views of atomic structure. Yellow dots are P atoms and the orange one is the K atom. The K atom is 3.5 Å away from the topmost P layer. (b) $k_x$-$k_y$ plot of calculated conduction and valence bands. (c) Calculated band structure along the $k_x$ axis (the zigzag direction). (d) Calculated band structure along the $k_y$ axis (the armchair direction).

## ASSOCIATED CONTENT

**Supporting Information**

Electronic structures of pristine bulk BP (Fig. S1), band structures of K-doped four-layer BP obtained from SIESTA and WIEN2k codes (Fig. S2), electronic wave functions of pristine four-layer BP (Fig. S3), K-density dependent band structures obtained from calculations using different supercell sizes (Fig. S4), K-density dependences of the band gap and the charge transfer (Fig. S5), and derivation of effective Hamiltonians from band dispersions. This material is available free of charge via the Internet at http://pubs.acs.org.

## AUTHOR INFORMATION

**Corresponding Author**

*E-mail: h.j.choi@yonsei.ac.kr

**ACKNOWLEDGEMENTS**

This work was supported by NRF of Korea (Grant No. 2011-0018306). K.S.K. acknowledges support from IBS-R014-D1. Y.Y. acknowledges support from NRF of Korea (Grant No. 2012M3A7B4049801), Samsung Display, and Yonsei University Future-leading Research Initiative of 2014 (2014-22-0123). Computational resources were provided by the KISTI Supercomputing Center (Project No. KSC-2014-C3-070).

**Supporting Information:**

**Emergence of Two-Dimensional Massless Dirac Fermions, Chiral Pseudospins, and Berry's Phase in Potassium Doped Few-Layer Black Phosphorus**

*Seung Su Baik,[1,2] Keun Su Kim,[3,4] Yeonjin Yi,[1] and Hyoung Joon Choi[1,2,]*

[1] Department of Physics, Yonsei University, Seoul 120-749, Korea.

[2] Center for Computational Studies of Advanced Electronic Material Properties, Yonsei University, Seoul 120-749, Korea.

[3] Department of Physics, Pohang University of Science and Technology, Pohang 790-784, Korea.

[4] Center for Artificial Low Dimensional Electronic Systems, Institute for Basic Science, Pohang 790-784, Korea.

*E-mail: h.j.choi@yonsei.ac.kr

# 1. Electronic structures of bulk BP

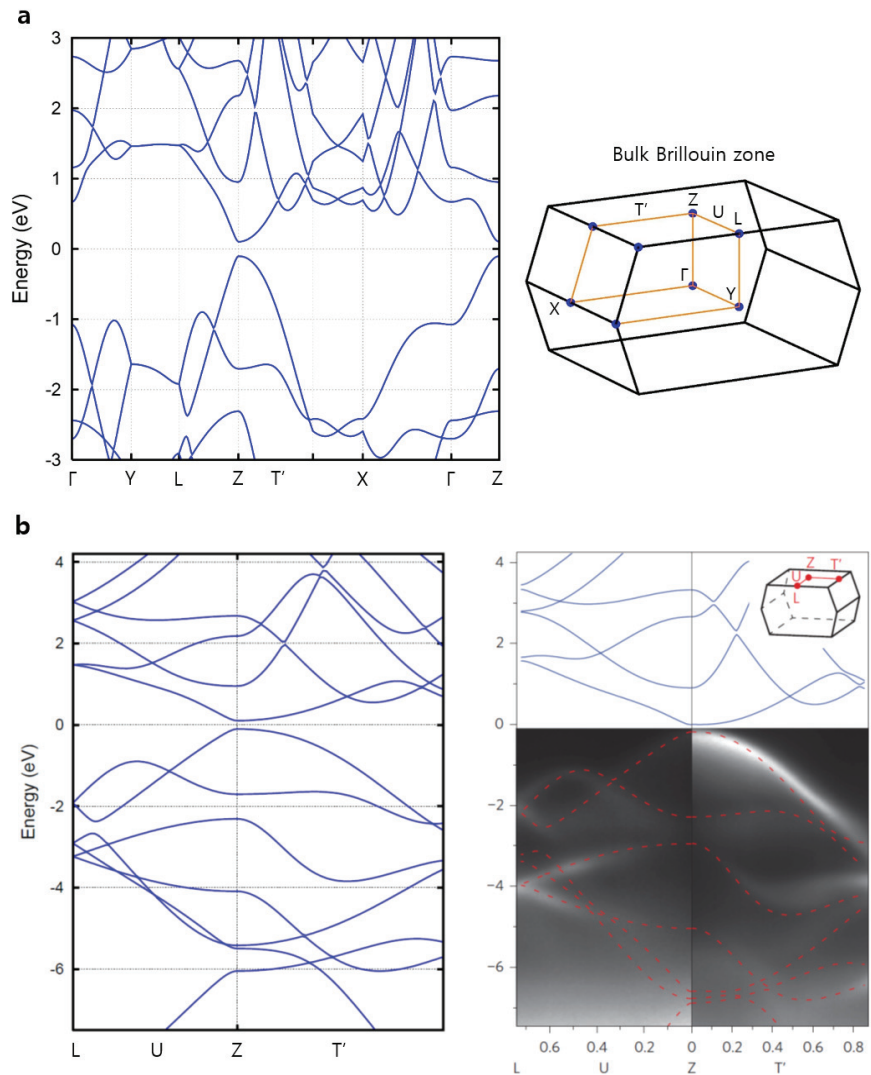

**Figure S1.** Electronic structure of bulk BP. (a) Band structure from GGA (PBE) calculation with relaxed cell parameters and atomic positions. (b) Comparison of our GGA result (Left) with the reported ARPES measurement (Right). In the right panel, superimposed blue solid and red dotted lines are the calculated bands by HSE06. Data was taken from Li et al, *Nat. Nanotechnol.* **2014**, *9* (5), 372-377.

**2. Band structure of K-doped four-layer BP: SIESTA versus WIEN2k**

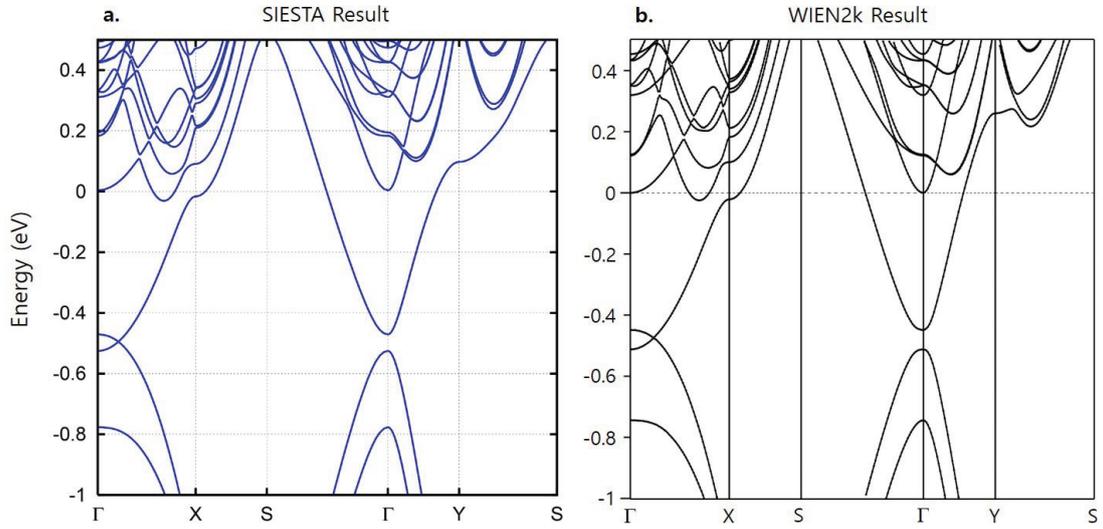

**Figure S2.** Band structures of K-doped four-layer BP, obtained from GGA (PBE) calculations. (a) SIESTA result for the K deposition on the 2×2 surface supercell. K atom and topmost BP layer were relaxed until the forces became less than 5.0 meV/Å. (b) WIEN2k result for the K deposition on the 2×2 surface supercell. K atom and topmost BP layer were relaxed until the forces became less than 0.2 mRy/a.u.

**3. Electronic wave functions of pristine four-layer BP**

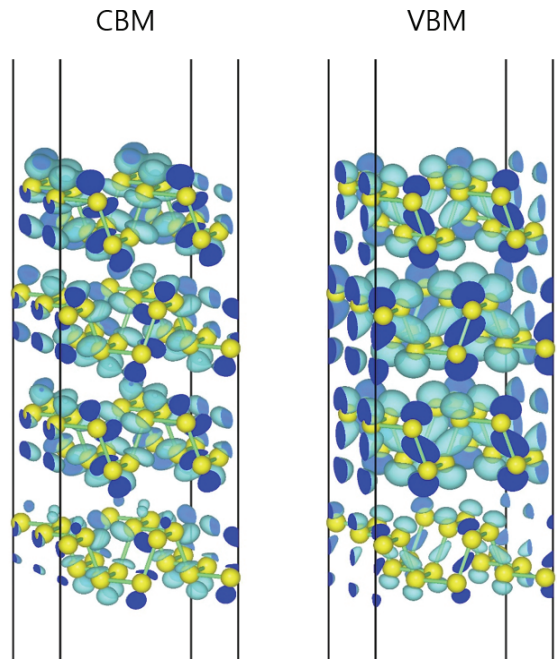

**Figure S3.** Wave functions in pristine four-layer BP. Electronic wave functions at CBM and VBM points. The isosurface level was set to $1.9\times10^{-3}/Å^3$.

## 4. Band gap versus surface K density: calculations with different supercells

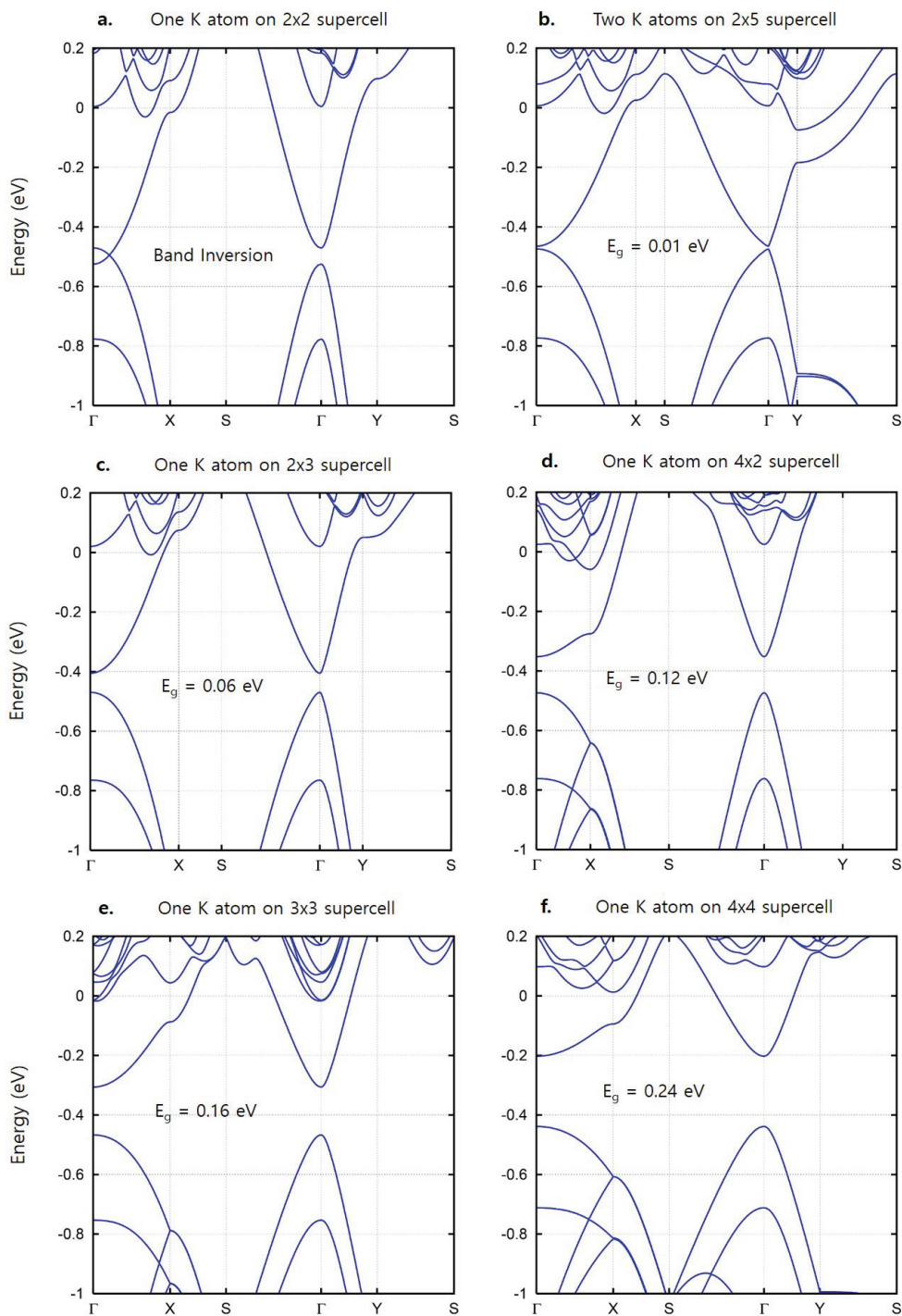

**Figure S4.** Band structures of K-doped four-layer BP. (a) One K atom on the $2\times2$ surface supercell (The surface K atom density is $n_K = 1.65\times10^{14}/cm^2$). (b) Two K atoms on the $2\times5$ surface supercell

($n_K = 1.32\times10^{14}$/cm$^2$). (c) One K atom on the 2×3 surface supercell ($n_K = 1.10\times10^{14}$/cm$^2$). (d) One K atom on the 4×2 surface supercell ($n_K = 8.23\times10^{13}$/cm$^2$). (e) One K atom on the 3×3 surface supercell ($n_K = 7.32\times10^{13}$/cm$^2$). (f) One K atom on the 4×4 surface supercell ($n_K = 4.12\times10^{13}$/cm$^2$). In all calculations, K atoms and topmost BP layers were relaxed until the forces became less than 5.0 $m$eV/Å.

## 5. Band gap versus surface K density, and charge transfer versus surface K density

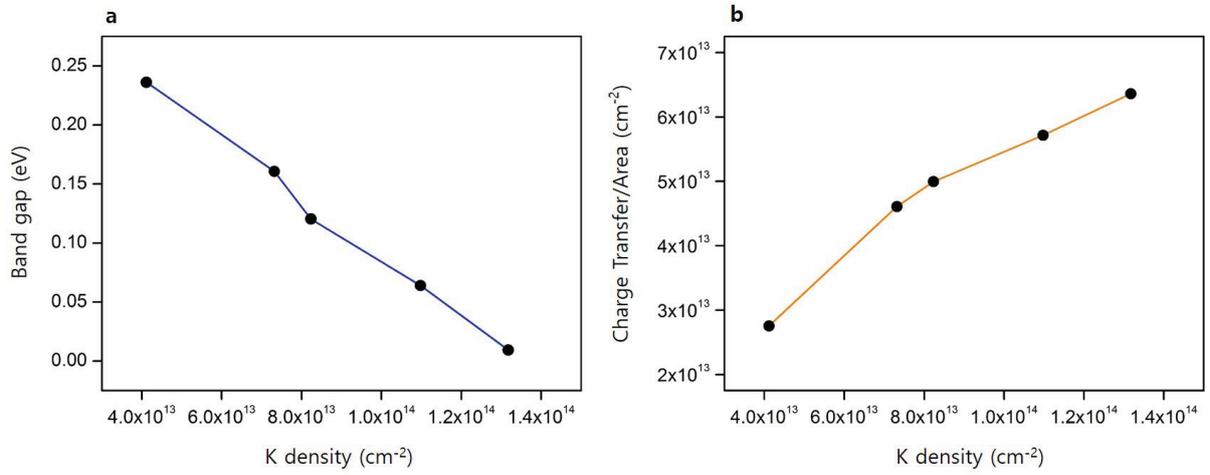

**Figure S5.** (a) Band gap as a function of the surface K atom density. (b) Charge transfer per surface area from K to BP layer as a function of the surface K atom density. (a) and (b) are obtained from the supercell calculations shown in Figs. S4(b)-(f). A linear fit of (a) yields a band-gap reduction rate of 0.28 eV per $1\times10^{14}$/cm$^2$.

## 6. Derivation of effective Hamiltonian

In the pristine four-layer PB without the potassium doping, the band dispersion along the $y$ direction

(the armchair direction) shows an asymptotic linear behavior as the k point goes away from the Γ point, as shown in Fig. 2d. This asymptotic linear dispersion is well fitted with a hyperbola, $E^2 = (E_g/2)^2 + (\hbar v_y k_y)^2$, including the Γ point. Here $E_g = E_c - E_v$ is the band gap. Thus, as a starting point, we can introduce a low-energy effective Hamiltonian,

$$H = \begin{pmatrix} E_c + a_x k_x^2 + a_y k_y^2 & \hbar v_y k_y \\ \hbar v_y k_y & E_v - b_x k_x^2 - b_y k_y^2 \end{pmatrix}$$

where $E_c$ is the conduction-band-minimum (CBM) energy, $E_v$ is the valence-band-maximum (VBM) energy, and $a_x$, $a_y$, $b_x$, $b_y$, and $v_y$ are constants. In this Hamiltonian, the eigenstate for the pseudospin $\sigma_z = +1$ is the CBM states and that for $\sigma_z = -1$ is the VBM state. Since we can set $(E_c + E_v)/2 = 0$ in general, we have $E_c = E_g/2$ and $E_v = -E_g/2$. Our band calculation results show that $a_x \approx b_x$, $a_y \approx 0$, and $b_y \approx 0$. Then, the Hamiltonian becomes

$$H = \begin{pmatrix} E_g/2 + a_x k_x^2 & \hbar v_y k_y \\ \hbar v_y k_y & -E_g/2 - a_x k_x^2 \end{pmatrix} = (E_g/2 + a_x k_x^2)\sigma_z + \hbar v_y k_y \sigma_x$$

For a better interpretation, we consider a matrix transformation, $H' = UHU^{-1}$, using

$$U = \frac{1}{\sqrt{2}} \begin{pmatrix} 1 & -i \\ 1 & i \end{pmatrix}.$$

Under this transformation, the identity matrix and the Pauli matrices are transformed as $UIU^{-1} = I$, $U\sigma_x U^{-1} = \sigma_y$, $U\sigma_y U^{-1} = \sigma_z$, and $U\sigma_z U^{-1} = \sigma_x$. Thus, after the transformation, Hamiltonian becomes

$$H = \begin{pmatrix} 0 & E_g/2 + a_x k_x^2 - i\hbar v_y k_y \\ E_g/2 + a_x k_x^2 + i\hbar v_y k_y & 0 \end{pmatrix} = (E_g/2 + a_x k_x^2)\sigma_x + \hbar v_y k_y \sigma_y.$$

Now, the eigenstate for the pseudospin $\sigma_x = +1$ is the CBM states and that for $\sigma_x = -1$ is the VBM state. The constant $a_x$ can be expressed using the effective mass ($m^*$), i.e., $a_x = \hbar^2/(2m^*)$.

With the K doping, the band gap starts to decrease but the asymptotic linear dispersion along the armchair direction is still retained while the slope of the asymptote is slightly deviated from its initial value. As the K-doping density is increased more, the band gap is reduced further. When the K doping reaches a critical value, making the conduction and valence bands touch at one k point as shown in Figs. 4c and d, we have $E_g = 0$ and the Hamiltonian becomes

$$H = \begin{pmatrix} 0 & a_x k_x^2 - i\hbar v_y k_y \\ a_x k_x^2 + i\hbar v_y k_y & 0 \end{pmatrix} = a_x k_x^2 \sigma_x + \hbar v_y k_y \sigma_y.$$

This Hamiltonian clearly shows that we now have massless Dirac Fermions along the $y$ direction (the armchair direction) and massive Dirac Fermions along the $x$ direction (the zigzag direction).

With the potassium deposition more than the critical density, $E_c$ is lower than $E_v$ ($E_c - E_v < 0$). Thus, when we define $k_D$ such that $E_g = -2a_x k_D^2 = -\hbar^2 k_D^2/m^*$, the Hamiltonian is expressed as

$$H = \begin{pmatrix} 0 & -a_x k_D^2 + a_x k_x^2 - i\hbar v_y k_y \\ -a_x k_D^2 + a_x k_x^2 + i\hbar v_y k_y & 0 \end{pmatrix} = (-a_x k_D^2 + a_x k_x^2)\sigma_x + \hbar v_y k_y \sigma_y,$$

and the conduction and valence bands cross at $\pm k_D$ along the $x$ direction. Near the cross point at $k_D$, the Hamiltonian can be expressed as

$$H = \begin{pmatrix} 0 & \hbar v_x (k_x - k_D) - i\hbar v_y k_y \\ \hbar v_x (k_x - k_D) + i\hbar v_y k_y & 0 \end{pmatrix} = \hbar v_x (k_x - k_D)\sigma_x + \hbar v_y k_y \sigma_y$$

where $v_x = 2a_x k_D/\hbar = \hbar k_D/m^*$ is the velocity along the $x$ direction at the Dirac point. This Hamiltonian describes the anisotropic massless Dirac Fermions in two dimensions.